\documentclass[prb,superscriptaddress,twocolumn,floatfix,amsmath,amssymb,showpacs,longbibliography]{revtex4-2}

\usepackage{multirow,booktabs}
\usepackage{graphicx}
\usepackage{float}
\usepackage{subfigure}
\usepackage{dcolumn}% Align table columns on decimal point
\usepackage{bm}% bold math
\usepackage{color}
\usepackage{diagbox}
\hfuzz=\maxdimen
\begin{document}

\title{Absence of itinerant ferromagnetism in a cobalt-based oxypnictide}
%\title{Sr$_{2}$CrCoAsO$_{3}$: A Layered Oxypnictide with Nearly Ideal CoAs$_4$ Tetrahedra}% Force line breaks with \\

\author{Hua-Xun Li}
\email[corresponding author: ]{lihuaxun@zju.edu.cn}
\affiliation{School of Physics, Interdisciplinary Center for Quantum Information, and State Key Laboratory of Silicon and Advanced Semiconductor Materials, Zhejiang University, Hangzhou 310058, China}

\author{Hao Jiang}
\affiliation{School of Physics and Optoelectronics, Xiangtan University, Xiangtan 411105, China}

\author{Yi-Qiang Lin}
\affiliation{School of Physics, Interdisciplinary Center for Quantum Information, and State Key Laboratory of Silicon and Advanced Semiconductor Materials, Zhejiang University, Hangzhou 310058, China}

\author{Jia-Xin Li}
\affiliation{School of Physics, Interdisciplinary Center for Quantum Information, and State Key Laboratory of Silicon and Advanced Semiconductor Materials, Zhejiang University, Hangzhou 310058, China}

\author{Shi-Jie Song}
\affiliation{School of Physics, Interdisciplinary Center for Quantum Information, and State Key Laboratory of Silicon and Advanced Semiconductor Materials, Zhejiang University, Hangzhou 310058, China}

\author{Qin-Qing Zhu}
\affiliation{School of Science, Westlake Institute for Advanced Study, Westlake University, Hangzhou 310064, China}

\author{Zhi Ren}
\affiliation{School of Science, Westlake Institute for Advanced Study, Westlake University, Hangzhou 310064, China}

\author{Guang-Han Cao}
\email[corresponding author: ]{ghcao@zju.edu.cn}
\affiliation{School of Physics, Interdisciplinary Center for Quantum Information, and State Key Laboratory of Silicon and Advanced Semiconductor Materials, Zhejiang University, Hangzhou 310058, China}
\affiliation{Collaborative Innovation Centre of Advanced Microstructures, Nanjing University, Nanjing, 210093, P. R. China}

\date{\today}% It is always \today, today,
\begin{abstract}

We report a layered transition-metal-ordered oxypnictide Sr$_{2}$CrCoAsO$_{3}$. The new material was synthesized by solid-state reactions under vacuum. It has an intergrowth structure with a perovskite-like Sr$_3$Cr$_2$O$_6$ unit and ThCr$_2$Si$_2$-type SrCo$_2$As$_2$ block stacking coherently along the crystallographic $c$ axis. The measurements of electrical resistivity, magnetic susceptibility, and specific heat indicate metallic conductivity from the CoAs layers and short-range antiferromagnetic ordering in the CrO$_{2}$ planes. No itinerant-electron ferromagnetism expected in CoAs layers is observed. This result, combined with the first-principles calculations and the previous reports of other CoAs-layer-based materials, suggests that the Co$-$Co bondlength plays a crucial role in the emergence of itinerant ferromagnetism.

\end{abstract}

\maketitle

\section{\label{sec:level1}Introduction}

The discovery of superconductivity in the iron-based pnictides~\cite{2008FeBased} stimulates the exploration of new superconductors with similar crystal structures in other transition-metal based systems. It is known that the key structural unit in iron-based superconductors is the antifluorite-like Fe$_2X_2$ ($X=$ As or Se) layer~\cite{2013JiangH,2015Hosono}.
While chemical doping at the Fe site with dopant Co or Ni induces superconductivity~\cite{2008PRB.1111.WangC,2008PRL.Sefat,2008PRB.1111Ni.GC}, a full replacement of Fe with other 3$d$-transition metals $M$ leads to distinct ground states including itinerant-electron antiferromagnetism ($M=$ Cr), antiferromagnetic insulators ($M=$ Mn), itinerant-electron ferromagnetic or enhanced Pauli-paramagnetic metals ($M=$ Co), conventional superconductors ($M=$ Ni), nonmagnetic metals ($M=$ Cu), and band insulators ($M=$ Zn)~\cite{2012PRB.SrCu2As2,2015Hosono,2020.122.IC}.

Those with CoAs layers, despite hosting no superconductivity, have attracted considerable attention because of diverse magnetic properties from nonmagnetic good metal (KCo$_2$As$_2$)~\cite{2022PRM.KCo2As2.Paglione,2022PRM.KCo2As2.Johnston}, renormalized/enhanced Pauli paramagnet (BaCo$_2$As$_2$)~\cite{2009PRB.Ba122Co.Sefat}, to itinerant-electron ferromagnetism (IEF) (LaCoAsO)~\cite{2008PRB.1111Co.Hosono}. A gapless quantum spin-liquid ground state~\cite{2013SrCo2As2} was suggested for SrCo$_2$As$_2$ where both ferromagnetic and striped antiferromagnetic spin fluctuations were detected~\cite{2013PRL.Sr122Co,2019PRL.Sr122Co}. Table~\ref{Co-based} summarizes the structural and physical properties of the CoAs-layer-based materials reported so far.
Interestingly, IEF was also observed in materials with enhanced two-dimensionality, such as Ba$_2$ScCoAsO$_3$~\cite{2017Ae2ScO3CoPn} and Sr$_2$ScCoAsO$_3$~\cite{2013Sr2ScO3CoAs}, with quite a large spacing between adjacent CoAs layers. The occurrence of IEF has been investigated in terms of structural properties~\cite{2017Ae2ScO3CoPn}. Nevertheless, the dominant factors governing the magnetic diversities remain elusive. Exploration of new materials containing CoAs layers will be helpful in unraveling this mystery.

\begin{table*}[htpb]
\center
\renewcommand{\arraystretch}{1.4}
\caption{Comparison of structural and physical properties of the compounds with CoAs layers. $h_\mathrm{As}$ denotes the As height from the Co plane. $\alpha$ is the As$-$Co$-$As bond angle along the $a$ or $b$ direction. $T_\mathrm{C}$ and $T_\mathrm{N}$ are temperatures of ferromagnetic and antiferromagnetic transitions associated with CoAs layers, respectively. IEF (IEAF) denotes itinerant-electron ferromagnetism (antiferromagnetism). N/A means that the data are not available, and "$-$" represents the case that is not applicable. Note that those materials containing magnetic rare-earth elements were not included for simplicity.}
\begin{tabular}{p{2.4cm}<{}p{1.2cm}<{}p{1.3cm}<{}p{1.4cm}<{}p{1.3cm}<{}p{1.3cm}<{}
p{1.7cm}<{}p{4.6cm}<{}p{1.4cm}<{}}
\hline
\hline
 Compounds & $a$ (\r{A}) & $c$ (\r{A}) & $d_\mathrm{Co-Co}$ (\r{A}) & $h_\mathrm{As}$ (\r{A}) & $\alpha$ ($^{\circ}$)   &  $T_\mathrm{C}$/$T_\mathrm{N}$ (K)     & Notes & Refs. \\\hline
  KCo$_2$As$_2$  & 3.794 & 13.570 & 2.683 & 1.506 & 103.1      & $-$       & Good metal        & \cite{1981KCo2As2,2022PRM.KCo2As2.Johnston} \\
  RbCo$_2$As$_2$ & 3.833 & 14.269  & 2.710 & 1.955 & 88.9       & $-$       & N/A                         & \cite{1984RbCo2As2} \\
  CsCo$_2$As$_2$ & 3.827 & 14.850  & 2.706 & 1.648 & 98.5       & $-$        & Pauli paramagnetic metal                       & \cite{1992CsCo2Pn2}\\
  CaCo$_2$As$_2$ & 3.989 & 10.330  & 2.821 & 1.147 & 120.2      & 76       & A-type IEAF (collapsed phase)      & \cite{2012CaCo2As2} \\
  SrCo$_2$As$_2$ & 3.947 & 11.773  & 2.791 & 1.280 & 114.3      & $-$       & FM/AFM spin fluctuations            & \cite{2013SrCo2As2} \\
  BaCo$_2$As$_2$ & 3.958 & 12.670  & 2.799 & 1.278 & 114.3      & $-$       & Enhanced Pauli paramagnetism      & \cite{2014BaCo2As2dopingK} \\
  LaCo$_2$As$_2$ & 4.054 & 10.324  & 2.866 & 1.414 & 110.2      & 86     & IEF metal (collapsed phase)    & \cite{2011LaCo2As2}\\
  \hline
  LiCoAs         & 3.758 & 6.170  & 2.657 & 1.481 & 103.5      & $-$        & Pauli paramagnetic metal       & \cite{1968LiCoAs,2018LiCoAs} \\
  \hline
  LaCoAsO        & 4.036 & 8.460  & 2.854 & 1.900 & 93.4       & 66       & IEF metal     & \cite{2008PRB.1111Co.Hosono,2008LaCoAsO} \\
  \hline
  Sr$_2$ScCoAsO$_3$ & 4.049 & 15.557  & 2.863 & 1.371 & 111.8   & 48       & IEF metal      & \cite{2013Sr2ScO3CoAs} \\
  Ba$_2$ScCoAsO$_3$ & 4.170 & 16.773  & 2.949 & N/A     & N/A       & 32       & IEF metal  & \cite{2017Ae2ScO3CoPn} \\
  Sr$_2$VCoAsO$_3$  & 3.940 & 15.489  & 2.786 & 1.332 & 111.9   & $-$       & No IEF, yet IEF with doping      & \cite{2018Sr2VCoAsO3} \\
  Sr$_2$CrCoAsO$_3$ & 3.913 & 15.571  & 2.767 & 1.378 & 109.7   & $-$        & Normal metal, absence of IEF       & This work \\
  \hline
  Sr$_2$CrCo$_2$As$_2$O$_2$ & $\sim$3.99  & $\sim$17.98 & $\sim$2.00 & N/A  & N/A   & $-$     & No IEF    & \cite{2017JPCS.21222Co}  \\
  \hline
  Sr$_3$Sc$_2$Co$_2$As$_2$O$_5$ & 4.074  & 26.471 & 2.881 & 1.183 & 119.7  & 41      & IEF metal& \cite{2019Sr3Sc2O5Co2As2} \\
\hline
\hline
\end{tabular}
\label{Co-based}
\end{table*}

Here we explore new CoAs-layer-based materials using the block-layer design strategy which has been efficient in finding Fe-based superconductors with intergrowth structures~\cite{2012Ba2Ti2Fe2As4O,2013JiangH,2021intergrowth}. According to the block-layer model, the interlayer charge transfer is essential and serves as a glue for the block-layer stacking. On the other hand, the lattice match between distinct block layers plays an important role as well, since it can minimize the elastic energy. Recently, we synthesized a Cr-based oxyarsenide Sr$_2$Cr$_2$AsO$_3$ consisting of alternating layers of ``Sr$_3$Cr$_2$O$_6$'' and ``SrCr$_2$As$_2$''~\cite{2022Sr2Cr2AsO3}. Note that the lattice parameter $a$ of SrCo$_2$As$_2$ (3.947 \AA) \cite{2013SrCo2As2} is close to that of SrCr$_2$As$_2$ (3.918 \AA) \cite{1980SrCr2As2}, satisfying lattice match with the Sr$_3$Cr$_2$O$_6$ block. Furthermore, Co$^{2+}$ ions in SrCo$_2$As$_2$ layers and Cr$^{3+}$ ions in Sr$_3$Cr$_2$O$_6$ layers have distinct valences, making the interlayer charge transfer possible. Thus, we designed the target material Sr$_{2}$CrCoAsO$_{3}$, which is isostructural to 21113-type Sr$_{2}$GaCuSO$_{3}$~\cite{1997Sr2GaCuSO3}, Sr$_{2}$CrCuSO$_{3}$~\cite{1997Sr2CrO3CuS}, and Sr$_{2}$VFeAsO$_{3}$~\cite{2009Sr2VO3FeAs}. We also checked the thermodynamic stability with theoretical calculations, indicating a negative hull energy of $-0.18$ eV per formula unit (f.u.).

In this paper, we report on the synthesis, crystal structure, and physical properties of the as-designed material Sr$_{2}$CrCoAsO$_{3}$. The structural analysis indicates that CoAs layers bear significant compressions from the relatively hard block layers of Sr$_{3}$Cr$_{2}$O$_{6}$.
The electrical resistivity data show a metallic behavior arising from the CoAs layers. The high-temperature magnetic susceptibility obeys the Curie-Weiss (CW) law, manifesting local-moment magnetism in the Sr$_{3}$Cr$_{2}$O$_{6}$ block layers. A short-range antiferromagnetic (AFM) ordering at $\sim$75 K is suggested from the magnetic and specific-heat data. No signature of CoAs-layer-associated IEF was detected. The absence of IEF is explained in terms of short Co$-$Co bondlength, associated with the Co-$d_{x^2-y^2}$-derived flat band away from the Fermi level.

\section{\label{sec:level2}EXPERIMENTAL METHODS AND DFT CALCULATIONS}
The polycrystalline sample of Sr$_{2}$CrCoAsO$_{3}$ was synthesized as follows using the source materials of Sr granules (99\%), As pieces (99.999\%), and powders of Cr (99.99\%), Cr$_{2}$O$_{3}$ (99.97\%), and Co$_{3}$O$_{4}$ (99.99\%). In the first step, the source materials were mixed in a nearly stoichiometric ratio (the oxygen content was reduced by 6\% to compensate for the slight oxidation of the reactive Sr). Then, the mixtures were put into an alumina crucible, reacted in an evacuated quartz ampoule at 800 $^{\circ}$C for 24~h and furnace-cooled to room temperature. After that, the sample was thoroughly ground for homogenization, pressed into a pellet, and sintered again in an evacuated silica tube at 1100$^{\circ}$C for 40 h. Finally, the sample was allowed to cool by stopping heating.

Powder X-ray diffraction (XRD) was carried out at room temperature on a PANalytical diffractometer utilizing Cu-\emph{K}$_{\alpha1}$ radiation. The structural refinement was performed using the program Rietan-FP \cite{2007rietveld}. The chemical compositions were confirmed by energy-dispersive X-ray spectroscopy (EDS, Oxford Instruments X-Max) equipped in a scanning electron microscope (SEM, Hitachi S-3700N).

Magnetic properties were measured on a Quantum Design magnetic property measurement system (MPMS3). Electrical resistivity and specific heat were measured on a Quantum Design Physical Property Measurement System (PPMS). For the resistivity measurement, a standard four-electrode method was employed. The specific-heat measurement employed a thermal relaxation method.

Density functional theory (DFT) calculations were done within the generalized gradient approximation (GGA)~\cite{1996BVS-1} by using the Vienna Ab-initio Simulation Package (VASP)~\cite{1996BVS-2}. A plane-wave basis energy cutoff of 550 eV was employed, along with a 15$\times$15$\times$4 $\Gamma$-centered k-mesh for calculations of density of states (DOS). We employed the experimental crystal structure parameters for the initial relaxation and obtained the relaxed structural parameters of $a_0=3.831$ \AA~ and $c_0=15.801$ \AA, which are respectively 2.1\% smaller and 1.5\% larger than the experimental values at room temperature. This relaxed structure was taken in the Bader valence charge calculations~\cite{2006BVS-3}. In order to study the Co$-$Co bondlength dependence of electronic structures, we also made structure optimizations by fixing the $a$ axis to 0.95$a_0$, 0.975$a_0$, 1.025$a_0$, and 1.05$a_0$, allowing other structural parameters to relax. The forces were minimized to less than 0.001 eV/\r{A} in the relaxation.

\
\section{\label{sec:level3}Results and discussion}
\begin{figure}[t]
\includegraphics[width=8cm]{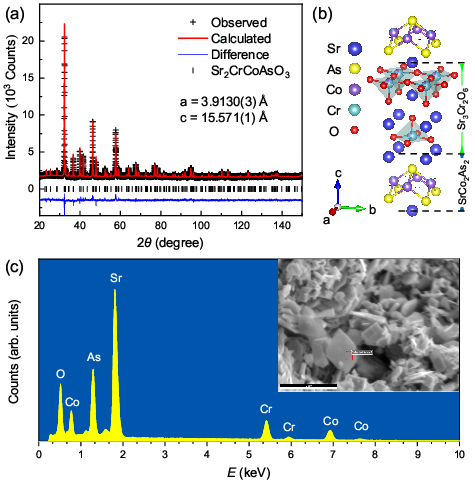}
\caption{(a) Room-temperature powder X-ray diffraction and its Rietveld refinement profile of Sr$_{2}$CrCoAsO$_{3}$. (b) The crystal structure consisting of alternating block layers of Sr$_{3}$Cr$_{2}$O$_{6}$ and SrCo$_{2}$As$_{2}$. (c) SEM image (inset) and typical EDS spectrum collected on the surface (marked by the small red rectangle) of a crystallite. The atomic ratios of Sr : Cr : Co : As : O are 1.99(8) : 1.00 : 0.90(5) : 1.05(5) : 3.2(3), basically consistent with the stoichiometry.
\label{XRD}
}
\end{figure}

Figure~\ref{XRD}(a) shows the XRD data of the Sr$_{2}$CrCoAsO$_{3}$ sample synthesized. The XRD pattern can be well indexed with the 21113-type lattice, suggesting the formation of the target compound. Thus we employed the structural model of Sr$_{2}$CrFeAsO$_{3}$ for the Rietveld analysis~\cite{2009Sr2CrO3FeAs1}. The structural refinement was very successful with a reasonably low reliable factor and the goodness of fit index $S$ (see Table~\ref{structure}). The refined lattice parameters are $a$ = 3.9130(3) {\AA} and $c$ = 15.571(1) {\AA}, both similar to those of the analogous compound Sr$_{2}$CrFeAsO$_{3}$ \cite{2009Sr2CrO3FeAs1,2009Sr2CrO3FeAs2}. As is the case in Sr$_{2}$CrFeAsO$_{3}$ \cite{2010eplSr2CrO3FeAs}, there is mutually mixed occupancy between Cr and Co, with about 11\% Co (Cr) occupying the Cr (Co) site. As shown in Fig.~\ref{XRD}(c), the chemical composition in a crystallite, determined by the EDS analysis, is consistent with the nominal one within the measurement uncertainty.

\begin{table}[t]
	\caption{ The crystallographic data of Sr$_{2}$CrCoAsO$_{3}$ obtained from Rietveld refinement of room-temperature powder XRD.}
	\label{structure}
	\small
	\scalebox{1.05}{
	\begin{tabular}{ccccccc}
		\hline \hline
		\multicolumn{3}{c}{Chemical Formula} & &  & \multicolumn{2}{c}{Sr$_{2}$CrCoAsO$_{3}$}\\
		\multicolumn{3}{c}{Crystal System} & & & \multicolumn{2}{c}{Tetragonal} \\
		\multicolumn{3}{c}{Space Group} & & & \multicolumn{2}{c}{$P$4/$nmm$ (No. 129)} \\
		\multicolumn{3}{c}{$a$ (\r{A})} & & & \multicolumn{2}{c}{3.9130(3)} \\
		\multicolumn{3}{c}{$c$ (\r{A})} & & & \multicolumn{2}{c}{15.571(1)} \\
		\multicolumn{3}{c}{$V$ (\r{A}$^{3}$)} & &  & \multicolumn{2}{c}{238.40(4)} \\
		\multicolumn{3}{c}{$Z$}          & & &    \multicolumn{2}{c}{2}  \\
\multicolumn{3}{c}{$R_{\mathrm{wp}}$}          & & &    \multicolumn{2}{c}{ 4.02\%}  \\
\multicolumn{3}{c}{$S$}          & & &    \multicolumn{2}{c}{1.72}  \\
		\hline
		Atoms       & site      & $x$      & $y$       & $z$         & Occupancy   & $U_{\mathrm{iso}}$ ({\AA}$^2$)   \\
		Sr(1)       & $2c$      & 0.00     & 0.50      & 0.1933(3)   & 1.000       & 0.0024  \\
		Sr(2)       & $2c$      & 0.00     & 0.50      & 0.4147(2)   & 1.000       & 0.0027  \\
		Cr(1)       & $2c$      & 0.50     & 0.00      & 0.3111      & 0.89       & 0.0126  \\
		Co(1)       & $2c$      & 0.50     & 0.00      & 0.3111(5)   & 0.11(1)   & 0.0127  \\
		Cr(2)       & $2a$      & 0.50     & 0.50      & 0.0000      & 0.11       & 0.0157  \\
		Co(2)       & $2a$      & 0.50     & 0.50      & 0.0000      & 0.89       & 0.0026  \\
		As          & $2c$      & 0.50     & 0.00      & 0.0885(3)   & 1.000       & 0.0068  \\
		O(1)        & $4f$      & 0.50     & 0.00      & 0.431(1)  & 1.000       & 0.0130  \\
		O(2)        & $2a$      & 0.50     & 0.50      & 0.2940(9)   & 1.000       & 0.0128  \\
		\hline
		\hline
	\end{tabular}
	}
\end{table}

The crystal structure, shown in Fig.~\ref{XRD}(b), consists of alternating antifluorite-type CoAs layers and perovskite-type Sr$_{3}$Cr$_{2}$O$_{6}$ blocks. The $a$ axis of the Sr$_{3}$Cr$_{2}$O$_{6}$ block is estimated to be 3.90 {\AA}~\cite{2022Sr2Cr2AsO3}. Thus, the lattice mismatch parameter, defined as $2|a_1-a_2|/(a_1+a_2)$ where $a_1$ and $a_2$ are respectively the $a$ axes of the two constituent blocks, is only 1.2\%, satisfying the lattice match constraint in the block layer structure model~\cite{2021intergrowth}. Note that the $a$ axis of the Sr$_{2}$CrCoAsO$_{3}$ (3.913 \AA) is closer to that of Sr$_{3}$Cr$_{2}$O$_{6}$ rather than SrCo$_{2}$As$_{2}$, indicating that the perovskite-like block layers have a larger value of stiffness. In other words, the CoAs layers bear compressive tensions, which leads to the shortest $a$ axis among all the 21113-type compounds listed in Table~\ref{Co-based}.

\begin{figure}
\includegraphics[width=8cm]{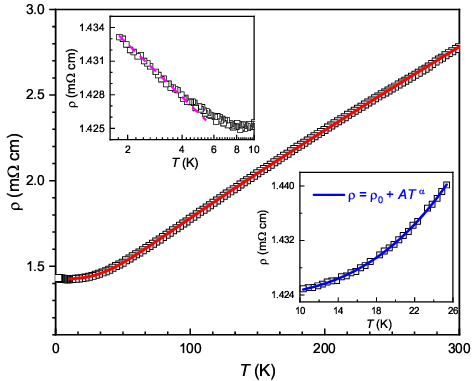}
\caption{Temperature dependence of electrical resistivity for the Sr$_{2}$CrCoAsO$_{3}$ polycrystalline sample. The solid line fits with an extended Bloch-Gr\"{u}neisen formula above 10 K. The top inset zooms in on the low-temperature resistivity data, and the bottom inset shows a fit with the power law.}
\label{RT}
\end{figure}

The temperature dependence of electrical resistivity for the Sr$_{2}$CrCoAsO$_{3}$ polycrystalline sample, $\rho(T)$, is shown in Fig.~\ref{RT}. The $\rho(T)$ curve exhibits a typical metallic behavior without discernible anomalies, akin to those of other CoAs-layer-based materials such as BaCo$_2$As$_2$~\cite{2009PRB.Ba122Co.Sefat}, SrCo$_2$As$_2$~\cite{2013SrCo2As2}, and LaCoAsO~\cite{2008PRB.1111Co.Hosono}. Since the analogous compound Sr$_{2}$CrCuSO$_{3}$ with identical Sr$_{3}$Cr$_{2}$O$_{6}$ layers is insulating~\cite{1997Sr2CrO3CuS}, one may conclude that the metallic conductivity in the title material arises from the CoAs layers.
At temperatures below 10 K, there is a slight resistivity upturn (top inset of Fig.~\ref{RT}) with log$T$ dependence, reminiscent of the Kondo effect \cite{1964KondoEffect}. As revealed by the XRD Rietveld analysis above, a small portion of Cr occupies the Co site in the CoAs layers, which should be the origin of the Kondo-like behavior. Note that an enhanced resistivity upturn at low temperatures was also observed in Sr$_2$CrFeAsO$_3$ with mixed Fe/Cr occupation~\cite{2009Sr2CrO3FeAs1,2010eplSr2CrO3FeAs}.

The $\rho(T)$ curve in the temperature range of 10 K $< T <$ 300 K can be well fitted by an extended Bloch-Gr$\mathrm{\ddot u}$neisen formula:
\begin{equation}
	\rho (T)  = \rho _{0}  + A(\frac{T}{\Theta _{R} }) ^{n} \int_{0}^{\Theta _{R}/T } \frac{x^{n} }{(e^{x}-1)(1-e^{-x})} dx,
\end{equation}
where $n$ is a parameter characterizing various conduction electron scattering types \cite{1959BGfit1,1964BGfit2,1960BGfit3}. The result yields $\rho _{0}$ = 1.42 m$\Omega$ cm, $\Theta _{R}$ = 265.1 K, and $n$ = 3.10. Indeed, the $\rho(T)$ data in the range of 10 K $< T <$ 25 K obey the power-law function, $\rho \left ( T \right )  = \rho _{0}  + A T^{\alpha } $, with $\alpha$ = 3.10. This $n$ or $\alpha$ value seems to suggest dominant magnetic scattering.

\begin{figure}
	\includegraphics[width=8cm]{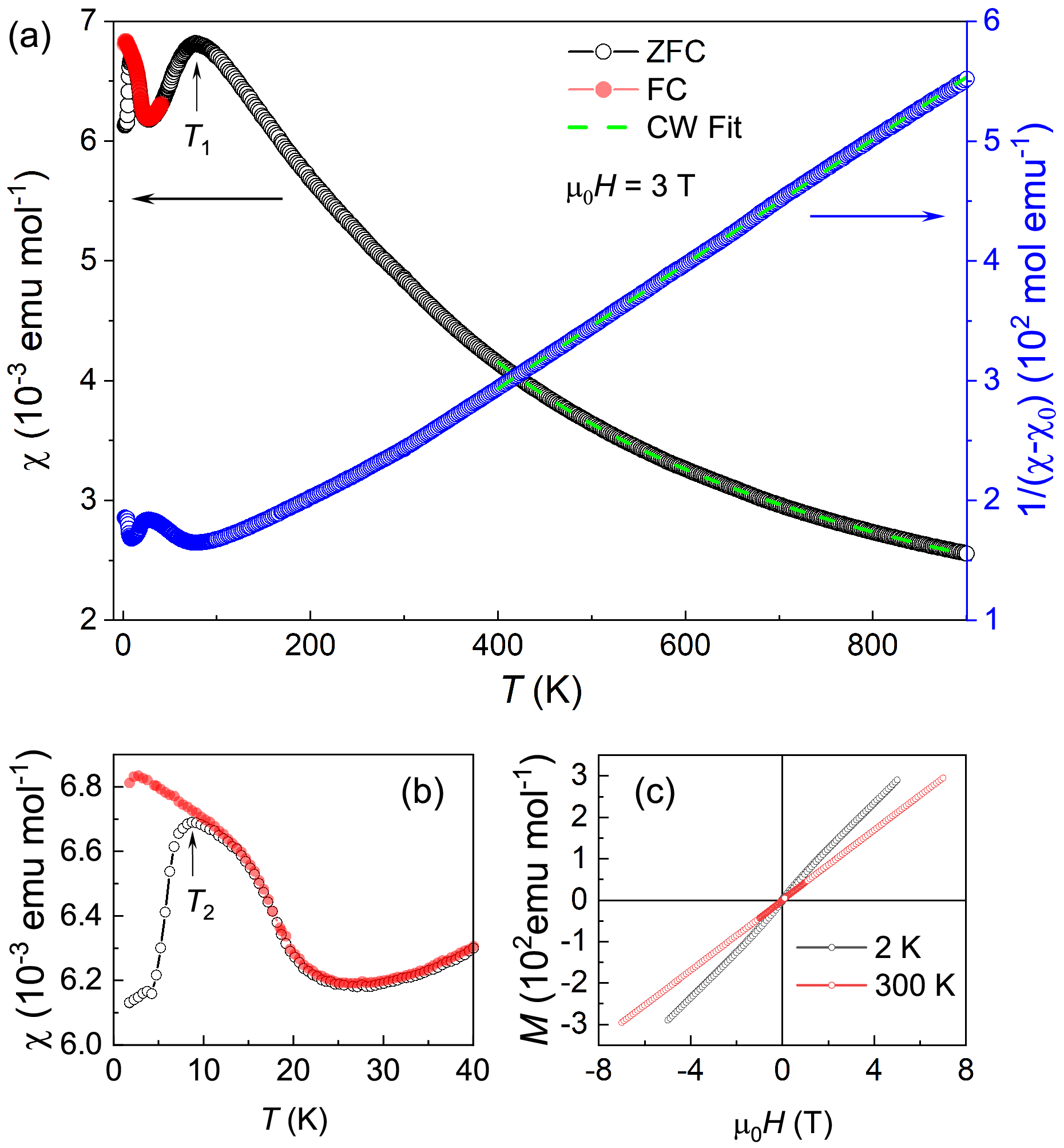}
    \caption{(a) Temperature dependence of magnetic susceptibility under a magnetic field of $\mu _{0}H$ = 3 T for Sr$_{2}$CrCoAsO$_{3}$. The right axis plots the reciprocal of susceptibility. The dashed lines show the CW fitting (see the text) in the temperature from 400 to 900 K. (b) A close-up in low temperatures. (c) Isothermal magnetization curves at 2 and 300 K.}
	\label{MT}
\end{figure}

Figure~\ref{MT} shows the magnetic measurement results for Sr$_{2}$CrCoAsO$_{3}$. The high-temperature $\chi(T)$ data follow a Curie-Weiss (CW) paramagnetic behavior. Fitting with the formula, $\chi \left ( T \right )  = \chi _{0}  + C/\left ( T + \theta _\mathrm{CW}  \right ) $, yields $\chi _{0} = 7.42 \times 10^{-4}$ emu mol$^{-1}$, $C$ = 1.92 emu K mol$^{-1}$, and $\theta _\mathrm{CW}$ = 164 K. With the fitted Curie constant $C$, the effective local magnetic moment is calculated to be $P _\mathrm{{eff}}$ = 3.92 $\mu _\mathrm{B}$/f.u., basically consistent with the expected value of high-spin Cr$^{3+}$ free ions (3.87 $\mu _\mathrm{B}$). The positive $\theta _\mathrm{CW}$ value suggests AFM correlations between Cr local moments. The result implies that the CoAs layers are Pauli paramagnetic with temperature-independent magnetic susceptibility.

In the low-temperature region, there are two obvious anomalies. One is the AFM-like hump at $T_\mathrm{1}\approx$ 75 K. A similar observation was reported for Sr$_{2}$CrCuSO$_{3}$ which contains the same block layers of Sr$_{3}$Cr$_{2}$O$_{6}$ ~\cite{1997Sr2CrO3CuS}. Nevertheless, it is not likely to be associated with long-range magnetic ordering because no specific-heat peak appears (see below). Instead, this susceptibility anomaly likely comes from short-range AFM ordering, as is the case in Sr$_{2}$Cr$_{2}$AsO$_{3}$~\cite{2022Sr2Cr2AsO3,2022IC.2213}. There exists another magnetic anomaly at $T_\mathrm{2}\approx$ 9 K where the FC and ZFC curves bifurcate [Fig.~\ref{MT}(b)]. A similar observation is also seen in Sr$_{2}$Cr$_{2}$AsO$_{3}$, which is attributed to Cr-spin reorientations, such as spin canting~\cite{2022Sr2Cr2AsO3}. Notably, no evidence of IEF can be detected, as is seen in the magnetization curve in Fig.~\ref{MT}(c).

\begin{figure}
\includegraphics[width=8cm]{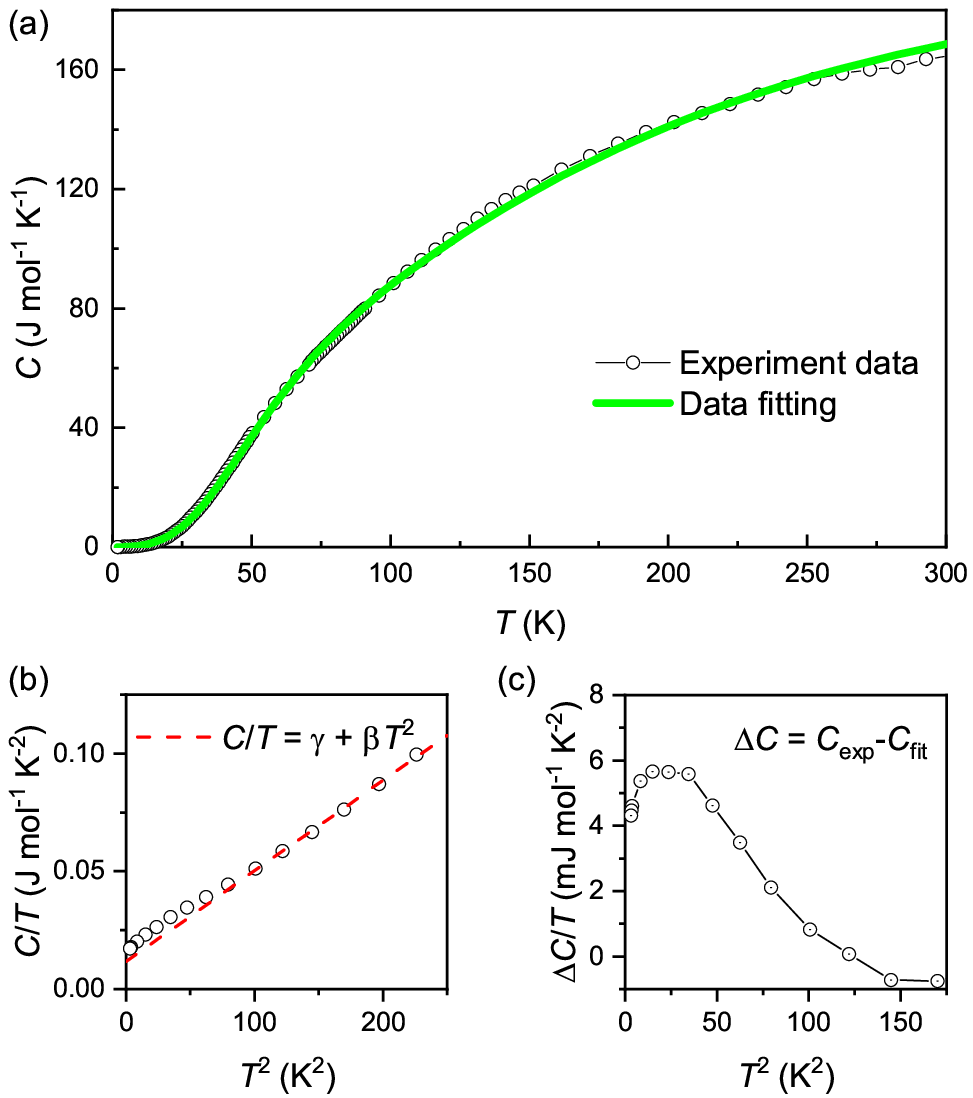}
\caption{(a) Temperature dependence of specific heat for Sr$_{2}$CrCoAsO$_{3}$.
         The green solid line is the fitted curve with a combined Debye and Einstein model (see the text). (b) The low-temperature specific-heat data, plotted with $C/T$ vs. $T^{2} $. The red dashed line fits with the formula $C\left ( T \right ) = \gamma T + \beta T^{3}$. The difference between experimental and fitted data is plotted in (c).}
\label{SH}
\end{figure}

Figure~\ref{SH} shows the temperature dependence of specific heat, $C(T)$, for the Sr$_{2}$CrCoAsO$_{3}$ sample. No obvious anomaly can be found in the whole temperature range. The result rules out the possibility of long-range magnetic ordering at $\sim$75 K. Instead, short-range magnetic ordering is possible, as in Sr$_{2}$Cr$_{2}$AsO$_{3}$~\cite{2022Sr2Cr2AsO3,2022IC.2213}. The absence of long-range order could be due to the two-dimensionality and/or the magnetic frustrations within the Sr$_{3}$Cr$_{2}$O$_{6}$ blocks. In the low-temperature regime, the specific heat from lattice contribution obeys Debye law, $C_\mathrm{lat}$ = $\beta T^{3}$. Therefore, the total specific heat can be ideally expressed as $C\left ( T \right ) = \gamma T + \beta T^{3}$. In the $C/T$ versus $T^2$ plot shown in Fig.~\ref{SH}(b), however, some deviations from the expected linearity can be seen below $\sim$10 K, coinciding with the slight resistivity upturn above, both in contrast to the observations in Sr$_{2}$Cr$_{2}$AsO$_{3}$~\cite{2022Sr2Cr2AsO3}. The linear fit in the temperature range from 10 to 15 K yields a Sommerfeld coefficient of $\gamma$ = 11.64 mJ K$^{-2}$ mol$^{-1}$. Figure~\ref{SH}(c) shows the difference between experimental and fitted data, which signifies a Kondo-like anomaly in heat capacity ~\cite{1984KondoExp,1987KondoExp,1981KondoCal,1982KondoCal}.

The whole $C(T)$ data can be well fitted by a combined formula with both Einstein and Debye components:
\begin{equation}
	\quad \quad  C(T) = {\xi}C_{\mathrm{D}}(\theta_{\mathrm{D}},T)+(1-{\xi})C_{\mathrm{E}}(\theta_{\mathrm{E}},T)+{\gamma}T,
	\label{eq1}
\end{equation}
where
\begin{displaymath}
	\quad \quad \quad C_{\mathrm{E}}(\theta_{\mathrm{E}},T)=3NR\left(\frac{\theta_{\mathrm{E}}/T}{e^{\theta_{\mathrm{E}}/T}-1}\right)^2e^{\theta_{\mathrm{E}}/T}
\end{displaymath}
and
\begin{displaymath}
	\quad \quad C_{\mathrm{D}}(\theta_{\mathrm{D}},T)=9NR\left(\frac{T}{\theta_{\mathrm{D}}}\right)^3\int_{0}^{\theta_{\mathrm{D}}/T}\frac{x^4e^{x}}{(e^{x}-1)^2}dx
\end{displaymath}
denote the lattice contributions in Einstein and Debye models, respectively. The coefficients $\xi$ and ($1-\xi$) are their individual proportions, which also represent the fractions of heavy (Sr, Cr, Co and As) and light (O) elements. With the $\gamma$ value fixed at 11.64 mJ K$^{-2}$ mol$^{-1}$, as a result, the remaining three parameters, namely $\theta_\mathrm{E}$, $\theta_\mathrm{D}$, and $\xi$, were fitted to be 723 K, 281 K, and 0.602, respectively. Note that the $\xi$ value is approximately 5/8, in line with the combined model. Also, the fitted Debye temperature is consistently close to the $\Theta_R$ value from the $\rho(T)$ data.

\section{\label{sec:level4}Discussion and Summary}

\begin{table}[h]
  \centering
  \caption{Bader charges of different elements and blocks in the magnetic ground state of Sr$_2$CrCoAsO$_3$.}
  \label{Bader}%
  \renewcommand{\arraystretch}{1.15}
  \scalebox{1.15}
    {
    \begin{tabular}{ccccc}
    \hline\hline
    Element & Charge & Neutral & Block & Block Charge \\
    \hline
    Sr(1) & 8.5855  & 10    & \multirow{4}[2]{*}{Sr$_3$Cr$_2$O$_6$} & \multirow{4}[2]{*}{-0.0119 } \\
    Sr(2) & 8.4717  & 10    &       &  \\
    Cr    & 10.4345  & 12    &       &  \\
    O     & 7.2660  & 6     &       &  \\
    \hline
    Sr(1) & 8.5855  & 10    & \multirow{3}[2]{*}{SrCo$_2$As$_2$} & \multirow{3}[2]{*}{0.0119 } \\
    Co    & 14.9363  & 15    &       &  \\
    As    & 15.7739  & 15    &       &  \\
    \hline\hline
    \end{tabular}%
	}
\end{table}

\begin{figure}[h]
\includegraphics[width=7.6cm]{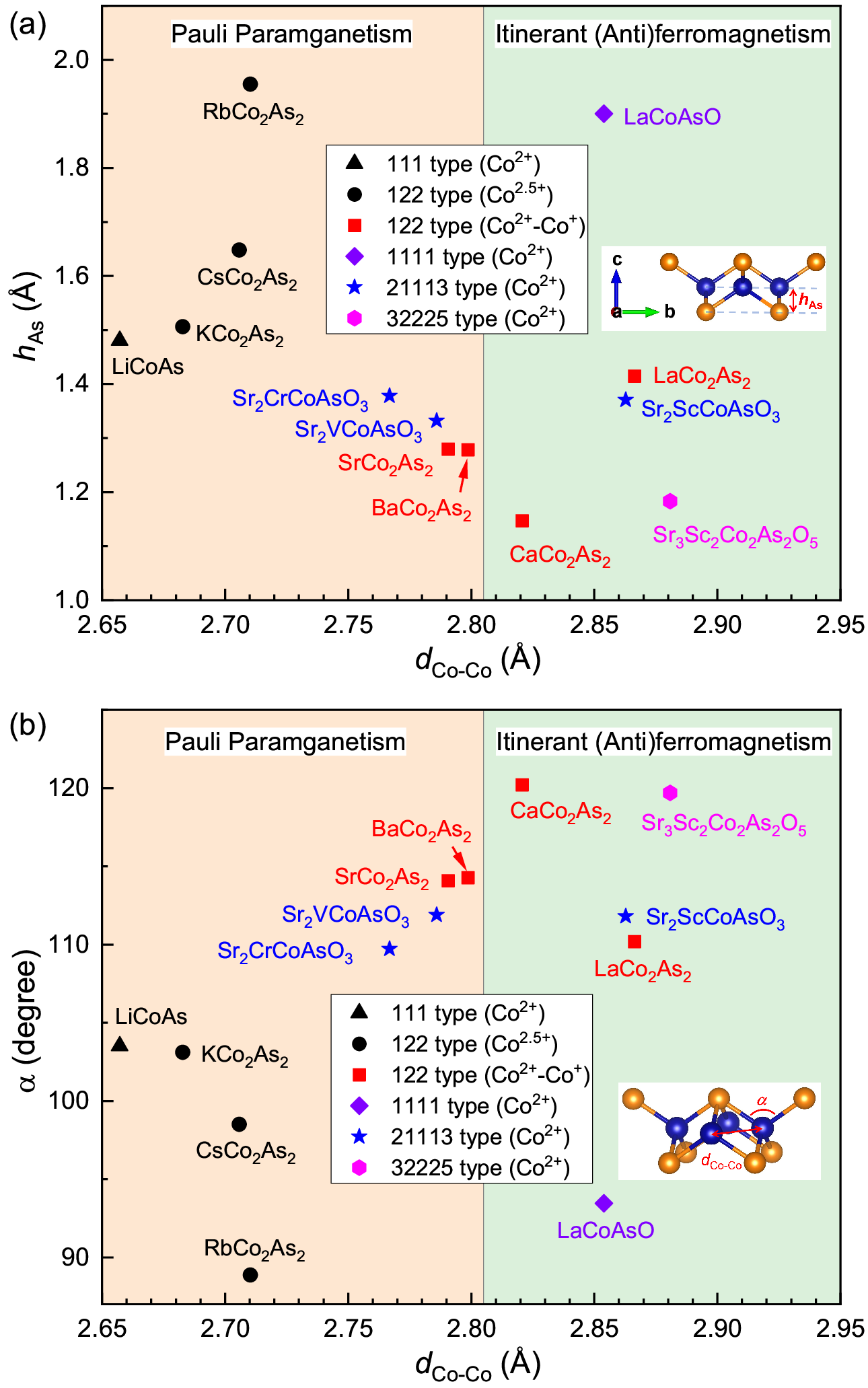}
\caption{Structure correlations with the emergent magnetism in CoAs-layer-based materials. The As height above the Co plane, $h_{\mathrm{As}}$ (a), and the As$-$Co$-$As bond angle, $\alpha$ (b), are plotted versus the Co$-$Co bondlength, respectively.}
\label{IEF-d-1}
\end{figure}

Above we demonstrated that the new CoAs-layer-based 21113-type Sr$_{2}$CrCoAsO$_{3}$ does not show IEF, which differs from other related materials such as Sr$_{2}$ScCoAsO$_{3}$ and Ba$_{2}$ScCoAsO$_{3}$ (see Table~\ref{Co-based}). For Sr$_{2}$VCoAsO$_{3}$, while it also does not exhibit IEF by itself, hole doping with Fe at the Co site leads to a ``revival'' of IEF~\cite{2018Sr2VCoAsO3}. In order to explore the possible self-electron doping due to inter-block-layer charge transfer in Sr$_{2}$CrCoAsO$_{3}$, we performed calculations of the Bader valence charges for Sr$_{2}$CrCoAsO$_{3}$. According to the results presented in Table~\ref{Bader}, the number of electrons transferred to the SrCo$_{2}$As$_{2}$ block is negligible. Thus any doping effect is unlikely to lead to the absence of IEF in the CoAs layers. Then questions are raised: What causes the absence of IEF in Sr$_{2}$CrCoAsO$_{3}$? Are there any structural parameters dominantly determining the emergence of IEF in CoAs-based compounds?

\begin{figure}[t]
\includegraphics[width=8.3cm]{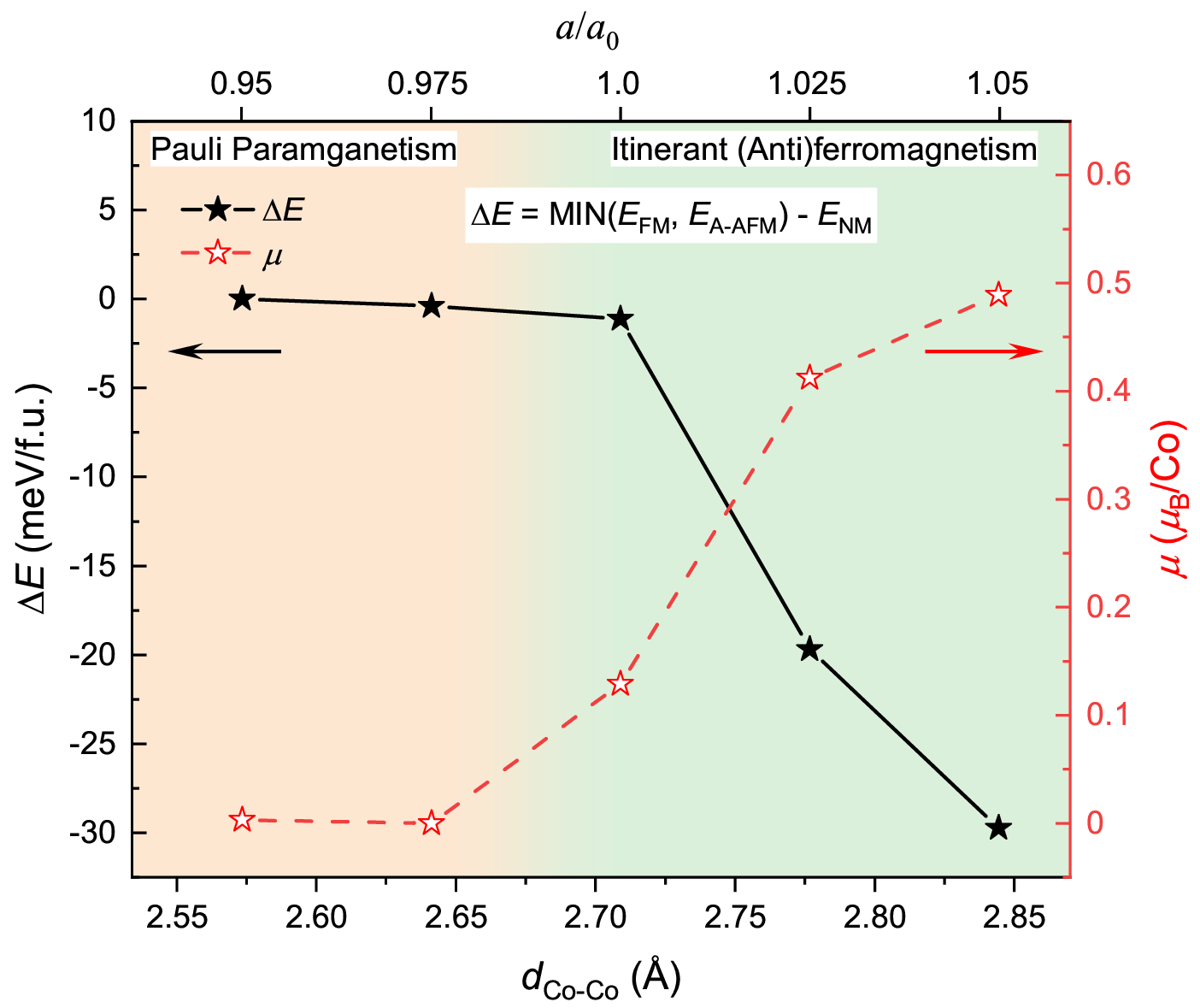}
\caption{The dependence of $\Delta$$E$ (left axis) and the effective magnetic moments of Co atoms (right axis) on $d_{\mathrm{Co}-\mathrm{Co}}$ in Sr$_{2}$CrCoAsO$_{3}$. MIN($E_\mathrm{FM}$, $E_\mathrm{A-AFM}$) takes the minimum energy between IEF and A-type IEAF states, while $\Delta$$E$ represents their energy relative to the NM state. The $a_0$ is the relaxed $a$-axis value based on experimental crystal structure parameters.}
\label{IEF-d-2}
\end{figure}

It is known that, in Fe-based superconductors, superconductivity is correlated with the geometry of FeAs$_4$ tetrahedra, the As height above the Fe plane ($h_{\mathrm{As}}$), and the As$-$Fe$-$As bond angle~\cite{2015Hosono,2010FeBasedReview}, both of which seem to signify the strength of the next-nearest-neighbor exchange interaction ($J_2$). Here we note that the nearest-neighbor exchange interaction ($J_1$) is generally related to $d_{\mathrm{Co}-\mathrm{Co}}$ owing to the direct exchange coupling, which should play a role for the magnetism. In Fig.~\ref{IEF-d-1}, therefore, we plot the $h_{\mathrm{As}}$ and the As$-$Co$-$As bond angle, $\alpha$, versus $d_{\mathrm{Co}-\mathrm{Co}}$ for the CoAs-layer-based systems. One immediately finds that IEF does not depend on either $h_{\mathrm{As}}$ or $\alpha$. By contrast, $d_{\mathrm{Co}-\mathrm{Co}}$ appears to be relevant to the magnetism. IEF appears in a distinct domain with $d_{\mathrm{Co}-\mathrm{Co}}>$ 2.8 \AA. On the other hand, Pauli paramagnetism dominates for the domain with $d_{\mathrm{Co}-\mathrm{Co}}<$ 2.8 \AA. At around the boundary of $d_{\mathrm{Co}-\mathrm{Co}}\approx$ 2.8 \AA, such as in the cases of BaCo$_2$As$_2$~\cite{2021BaCo2As2dopingIr}, SrCo$_2$As$_2$~\cite{2019PRB.122NFL}, and even Sr$_{2}$VCoAsO$_{3}$~\cite{2018Sr2VCoAsO3}, IEF or short-range ferromagnetic order can be recovered by the chemical doping that substantially increases the $a$ axis. It is noted that the title material Sr$_{2}$CrCoAsO$_{3}$ is located at the Pauli-paramagnetic domain. The reduced $a$ axis generally increases the bandwidth of Co-3$d$-related bands and, consequently, DOS at Fermi energy decreases. This may lead to the failure of the Stoner criterion for IEF~\cite{1938.Stoner}. This is probably the main reason for the absence of IEF in Sr$_{2}$CrCoAsO$_{3}$.

\begin{figure}[t]
\includegraphics[width=8.41cm]{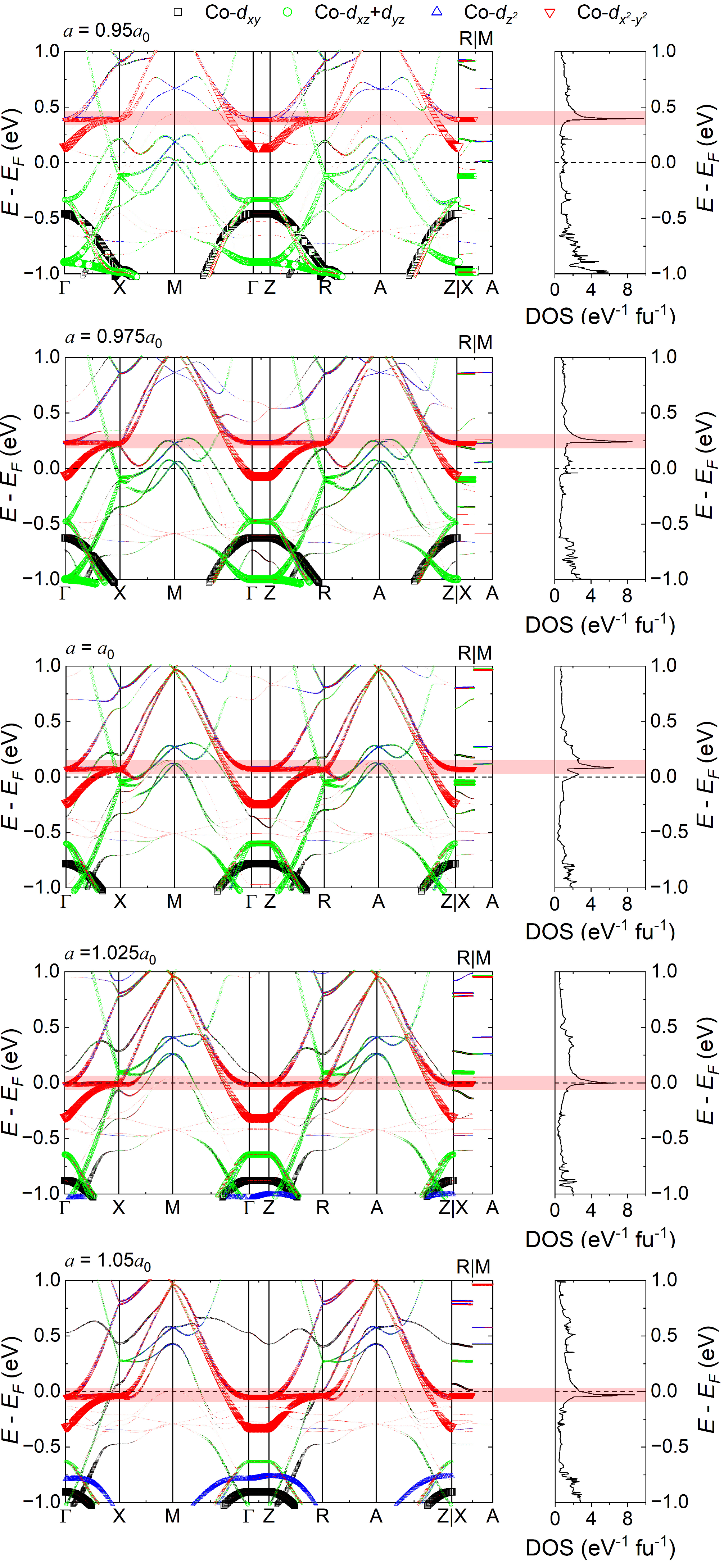}
\caption{Calculated electronic structures of Co-3$d$ electrons and corresponding DOS in Sr$_{2}$CrCoAsO$_{3}$ with the relaxed structures. The red shadows highlight the flat bands contributed by Co$-d_{x^2-y^2}$ orbitals. The marker size represents the spectral weight.}
\label{BandStructure}
\end{figure}

In order to further elucidate the crucial role of $d_{\mathrm{Co}-\mathrm{Co}}$ in determining the magnetism of the CoAs-layer systems, we performed DFT calculations for Sr$_{2}$CrCoAsO$_{3}$ assuming a tensile or compressive strain within the $ab$ plane. As shown in Fig.~\ref{IEF-d-2}, under compression, the ground state of CoAs layers for $a<0.975a_0$ are nonmagnetic (or Pauli paramagnetic). By contrast, with a tensile strain of $a>a_0$, the IEF state is stabilized, accompanied by a synchronous increase in the ordered magnetic moments. Spin polarizations in the IEF states are displayed in Fig. S1 of the Supplemental Material~\cite{SM}. The result directly verifies the effectiveness of $d_{\mathrm{Co}-\mathrm{Co}}$ as a parameter for IEF in the CoAs-layer-based system.

Figure~\ref{BandStructure} shows the electronic DOS as well as the band structures in nonmagnetic states highlighting the contributions from Co-3$d$ orbitals. There is a DOS peak associating with the flat band mainly contributed by the Co$-d_{x^2-y^2}$ orbital. With the increase of $a$ axis (or, $d_{\mathrm{Co}-\mathrm{Co}}$), the flat band moves progressively to the Fermi level. In addition, the Fermi level touches the flat band at $a$ = 1.025$a_0$, maximizing DOS at $E_\mathrm{F}$, in line with the Stoner criterion. Note that this phenomenon was also observed in CaCo$_{2}$As$_{2}$ with IEAF~\cite{2018SrCo2As2}.
%Based on the above calculation results and the law summarized from Fig.~\ref{Summary}, we conclude that $d_{\mathrm{Co}-\mathrm{Co}}$ is closely intertwined with IEF in CoAs-based materials. A large $d_{\mathrm{Co}-\mathrm{Co}}$ can make the Fermi level close to the van Hove singularity, where high DOS promotes ferromagnetic low-energy spin excitations and thus materials exhibit IEF according to the Stoner criterion. Besides, we also studied magnetic ground states and DOS of Sr$_{2}$CrCoAsO$_{3}$ based on the same relaxed structures above. As plotted in Fig. S1 and S2, with the increase of $d_{\mathrm{Co}-\mathrm{Co}}$, the compound magnetic ground state transitions from nonmagnetism to FM/A-type AFM, accompanied by a synchronous increase in the effective magnetic moments of Co atoms, and the DOS of spin up and spin down begin to stagger gradually denoting the emergence of IEF characteristics. Therefore, the calculation results of considering nonmagnetic state and (anti)ferromagnetic state are consistent.}

In summary, we successfully synthesized a Co-based oxypnictide Sr$_{2}$CrCoAsO$_{3}$, which has an intergrowth structure with alternating SrCo$_{2}$As$_{2}$ and Sr$_{3}$Cr$_{2}$O$_{6}$ layers. It was found that the SrCo$_{2}$As$_{2}$ block layers are responsible for the metallic behavior, while the Sr$_{3}$Cr$_{2}$O$_{6}$ slab mainly accounts for the CW paramagnetism at high temperatures, followed by short-range AFM ordering below $\sim$75 K. No IEF behavior from the CoAs layers was detectable.
%\textcolor{red}{We find that the Co$-$Co bondlength is a simple parameter that dominantly determines the magnetic properties of CoAs-based compounds.}
The absence of IEF in the title material is explained in terms of the compression of the Co$-$Co bondlength, which pushes the Co$-d_{x^2-y^2}$-derived flat band away from the Fermi level, and ultimately falls short of the Stoner criterion for IEF.

\begin{acknowledgments}
This work was supported by the National Natural Science Foundation of China (Grant No. 12005003), the National Key Research and Development Program of China (Grant No. 2022YFA1403202), and the Key Research and Development Program of Zhejiang Province, China (Grant No. 2021C01002).
\end{acknowledgments}

%\bibliography{CoAs21113}


\end{document}